%% file: main.tex
\documentclass{ieeeaccess}
\usepackage{cite}
\usepackage{amsmath,amssymb,amsfonts}
\usepackage{algorithmic}
\usepackage{graphicx}
\usepackage{textcomp}

\usepackage{threeparttable}
\usepackage{tabularx,booktabs}
\usepackage{diagbox}
\newcolumntype{C}{>{\centering\arraybackslash}X} 
\usepackage{xcolor}
\usepackage{pict2e}

\newsavebox{\ORCIDlogo}
\savebox{\ORCIDlogo}{%
\setlength{\unitlength}{\dimexpr 1em/256\relax}%
\begin{picture}(256,256)%
  \color[HTML]{A6CE39}\put(128,128){\circle*{256}}%
  \color{white}%
  \put(78.6,199.2){\circle*{20}}%
  \moveto(70.9,176,9)\lineto(86.3,176,9)\lineto(86.3,69.8)\lineto(70.9,69.8)%
  \closepath\fillpath%
  \moveto(108.9,176.9)\lineto(150.5,176.9)%
  \curveto(190.1,176.9)(207.5,148.6)(207.5 ,123.3)%
  \curveto(207.5,95,8)(186,69.7)(150.7,69.7)%
  \lineto(108.9,69.7)%
  \closepath\fillpath%
  \color[HTML]{A6CE39}%
  \moveto(124.3,83.6)\lineto(148.8,83.6)%
  \curveto(183.7,83.6)(191.7,110.1)(191.7,123.3)%
  \curveto(191.7,144.8)(178,163)(148,163)%
  \lineto(124.3,163)%
  \closepath\fillpath%
\end{picture}%
}

\newcommand\orcidicon[1]{\href{https://orcid.org/#1}{\usebox{\ORCIDlogo}}}

\usepackage{hyperref} 
\hypersetup{
    colorlinks=true,
    allcolors=blue,
}

\def\BibTeX{{\rm B\kern-.05em{\sc i\kern-.025em b}\kern-.08em
    T\kern-.1667em\lower.7ex\hbox{E}\kern-.125emX}}
\begin{document}
\history{Received June 2, 2023, accepted July 17, 2023. Date of publication xxxx 00, 0000, date of current version xxxx 00, 0000.}
\doi{10.1109/ACCESS.2023.0322000}

\title{SoK: A Taxonomy for Critical Analysis of Consensus Mechanisms in Consortium Blockchain}
\author{\uppercase{Wei Yao}\orcidicon{0000-0001-5019-3216}\authorrefmark{1}, \IEEEmembership{Student member, IEEE},
\uppercase{Fadi P. Deek}\authorrefmark{1}, \uppercase{Renita Murimi}\orcidicon{0000-0003-0674-6228}\authorrefmark{2}, \IEEEmembership{Senior member, IEEE}, and Guiling Wang\orcidicon{0000-0003-1880-4763}\authorrefmark{1},
\IEEEmembership{Fellow, IEEE}}

\address[1]{New Jersey Institute of Technology, University Heights, Newark, NJ 07102 USA}
\address[2]{University of Dallas, 1845 East Northgate Drive
Irving, TX 75062 USA}
\tfootnote{The research is partially supported by FHWA EAR 693JJ320C000021}

\markboth
{Wei \headeretal: SoK: A Taxonomy for Critical Analysis of Consensus Mechanisms in Consortium Blockchain}
{Wei \headeretal: SoK: A Taxonomy for Critical Analysis of Consensus Mechanisms in Consortium Blockchain}

\corresp{Corresponding author: Wei Yao\orcidicon{0000-0001-5019-3216} (e-mail: wy95@njit.edu)}

\begin{abstract}
\input{0_abstract}
\end{abstract}


\titlepgskip=-21pt

\maketitle

\input{1_intro}
\input{2_overview}
\input{3_taxonomy}
\input{4_reliability}
\input{5_perf}

\input{6_security}

\input{7_conclusion}

\section*{Acknowledgment}
The research is partially supported by FHWA EAR 693JJ320C000021.

\bibliographystyle{ieeetr}
\bibliography{reference}

\EOD

\end{document}

%% file: 0_abstract.tex
Consensus algorithms are central to blockchain technology and an emerging research area. In this paper, we begin with an overview of the different types and architectures of blockchain networks. Then, with a focus on consortium blockchains, we survey, classify, and assess their principal consensus mechanisms. 
Furthermore, as consensus mechanisms determine network reliability, enhance performance efficiency, and ensure system security, we conduct a critical analysis of the strengths and weaknesses of consensus algorithms using a taxonomy of three different criteria: reliability, performance, and security. 
We conclude with insights into current and future research challenges and opportunities in this domain.

%% file: 1_intro.tex
\section{Introduction}
Blockchain, a decentralized, immutable, and transparent distributed ledger, maintains a continuously growing list of transaction records ordered into blocks. At its core lies the so-called consensus algorithm, an agreement to validate the correctness of blockchain transactions. By their nature, public blockchains are resource-intensive technology. For example, in Bitcoin, each node uses the Proof of Work (PoW) algorithm to reach a consensus by competing to solve a puzzle \cite{nakamoto_bitcoin_2009}. Being less resource-intensive, a private blockchain functions as a distributed but secure ledger for a particular organizational purpose. A hybrid of both public and private blockchains, a consortium blockchain is an enterprise-level ledger that does not contend with issues of creating a resource-saving global consensus protocol like public chains. This section covers some important fundamental aspects of blockchain and overviews related work.

Nakamoto proposed Bitcoin in 2008 \cite{nakamoto_bitcoin_2009}, heightening interest in the blockchain technology that forms the foundation for digital currency \cite{Swan_2015}. The consensus algorithm provides a process for all nodes to seek and reach a common agreement in a distributed, increasingly untrusted environment.
Since Bitcoin, many cryptocurrencies have emerged \cite{dotan2020sok}. Among them, Ethereum \cite{wood_ethereum_nodate} is noteworthy for introducing the concept of smart contracts, which allow contracts to be coded on the blockchain and use Ethereum as a platform for currency transactions. Ethereum and Bitcoin share the feature of being public and allowing any node to participate in network activities, with similar consensus mechanisms.
In 2015, the Linux Foundation initiated Hyperledger Fabric \cite{Androulaki_Barger} as a solution specifically designed for enterprise-level applications, in contrast to the open nature of Bitcoin and Ethereum without any authentication mechanisms. Unlike Bitcoin's incentive mechanism, Fabric uses permissioned blockchain to increase energy efficiency and performance. With the rapid development of blockchain technology, more enterprise-level users have begun considering blockchain to meet their business needs \cite{hyperledger_walmart,hyperledger_culedger,hyperledger_kubernetes}. 
Therefore, exploring effective consensus protocols for use in consortium blockchains has developed into a research problem of emerging significance. 
The release of Facebook’s Libra project in 2019 \cite{libra_association_members_libra_2020} led to a new round of cryptocurrency interest, which in turn further increased attention from investors and researchers. Among various applications of blockchain, a notable one is that of digital governance. In what is touted as Web 3.0, the prevalent use of blockchain has accelerated the pace of innovation; thus, the requirements for consensus have also risen to a new level.

There are a number of surveys on blockchain and applications \cite{publicblockchainsurvey,Thakur_Kulkarni_2017, casino2019systematic, zou2020focus}. Similarly, surveys on blockchain consensus protocols have been published in the literature \cite{Consortium, Nguyen_Survey} and presented on arXiv \cite{Salimitari_Chatterjee_2019, Xiao_Zhang_Lou_Hou_2019, Cachin2017BlockchainCP}. Nguyen et al. \cite{Nguyen_Survey} provided a tutorial-style review on distributed consensus protocols that classifies consensus algorithms into proof-based and voting-based on the mechanism of reaching consensus. Important protocols, such as RBFT, HotStuff, and LibraBFT are not covered. Salimitari et al. \cite{Salimitari_Chatterjee_2019} reported on consensus algorithms and their applicability in the IoT areas. As noted \cite{Nguyen_Survey}, multiple important protocols, including LibraBFT, are missing. Consensus protocols such as LibraBFT \cite{libra_association_members_libra_2020} are relevant not only because they are suitable for enterprise scenarios but because they include features of public blockchain consensus protocols, such as incentive mechanisms. The survey of Ferdous et al. \cite{Ferdous2020BlockchainCA} also missed multiple important protocols. Recently, Badr et al. \cite{bellaj2022sok} provided a comprehensive survey on distributed ledger technologies, but they mainly focused on the network models, not consensus algorithms.

Given that a comprehensive survey covering all the important consensus protocols for consortium blockchains remains missing, as well as considering the importance of consensus mechanisms and the rapid development of enterprise-level blockchains, this paper fills a gap by providing a systematic taxonomy of enterprise-level blockchain consensus protocols and a detailed analysis of each protocol, considering aspects such as reliability, performance, and security.
The remaining parts of this paper are organized as follows: Section \ref{sec:BG} overviews blockchain technology and the notion of consensus algorithms. Section \ref{sec:taxonomy} presents our proposed taxonomy with algorithmic workflow for achieving fault tolerance and a comparative analysis of representative categories. Section \ref{sec:reliability} addresses the degree of achieved reliability. Section \ref{sec:performance} focuses on performance, while section \ref{sec:security} focuses on security. Finally, Section \ref{sec:conclusion} offers concluding remarks and presents research challenges and opportunities to motivate future work.


%% file: 2_overview.tex
\section{Building Blocks} \label{sec:BG}
\subsection{Blockchain Overview}
\begin{figure}[htp]
    \centering
    \includegraphics[width=\linewidth]{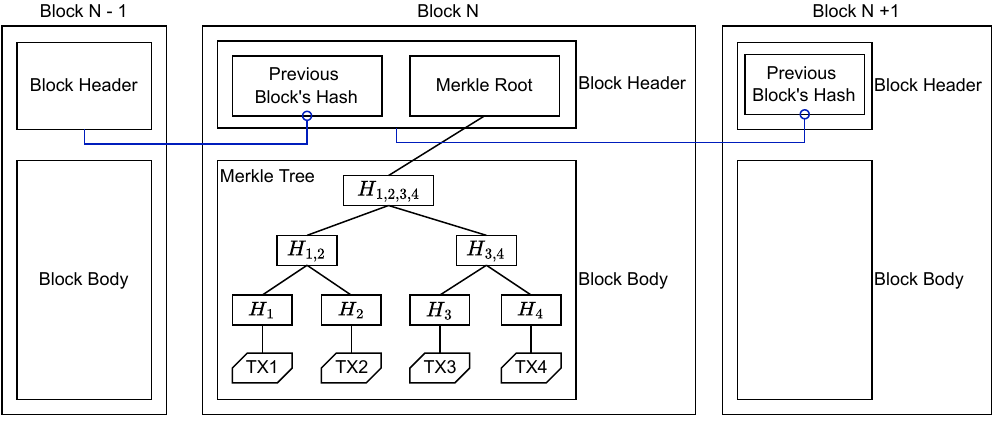}
    \caption{An example of block and chain structure}
    \label{fig:block}
\end{figure}

The nomenclature of \textbf{blockchain} is derived from its architecture: each block is linked cryptographically to the previous block. The \textbf{genesis block} is the first block, and each block has a set of transactions. Blockchain has the following characteristics: decentralization, trustlessness, immutability, and anonymity. Decentralization means there's no central trusted third party. Trust is established through consensus. Blockchain is tamper-proof, and it ensures some degree of anonymity and privacy-protection technologies like group signatures, ring signatures, and zero-knowledge proofs \cite{Bernal}.
\begin{figure}[htbp]
    \centering
    \includegraphics[width=\linewidth]{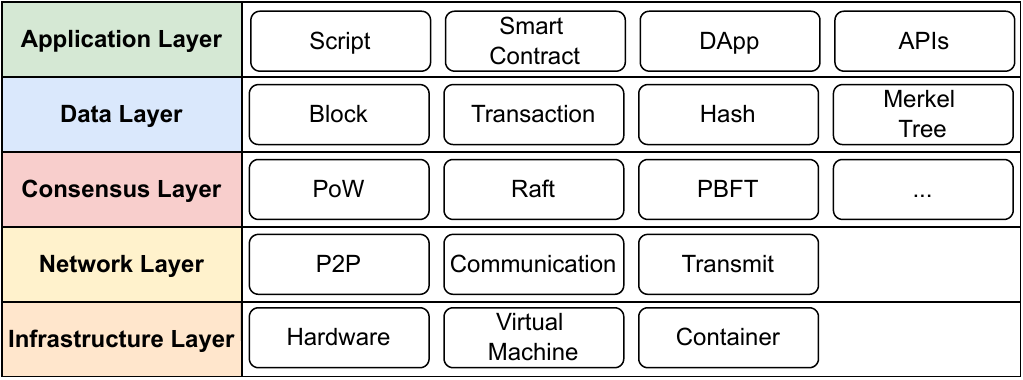}
    \caption{Blockchain Architecture}
    \label{fig:archi}
\end{figure}
The framework of the blockchain is shown in Figure \ref{fig:archi}. It comprises the following layers from bottom up: (1) \textbf{Infrastructure Layer}: Contains hardware, architecture equipment, and deployment environment for blockchain systems. 
(2) \textbf{Network Layer}: Includes the node organization method, communication mechanisms, and transmit protocol. P2P networks are used with a flat topology, where nodes are equal, distributed, and autonomous \cite{Mao_Deb_Venkatakrishnan_Kannan_Srinivasan_2020}.
(3) \textbf{Consensus Layer}: Forms the core of the consensus protocol used to ensure trust and security in the network and to achieve consistency of nodes participating in the distributed ledger.
(4) \textbf{Data Layer}: To realize traceability and non-tampering, transaction data is recorded through the blockchain structure and can be tracked through this chain ledger \cite{Paik_Xu_Bandara_Lee_Lo_2019}. For example, each data block in Bitcoin comprises a block header and a block body containing a packaged transaction, shown in Figure \ref{fig:block}. The block header contains information such as the current system version number, the previous block's hash value, the random number, the root of the Merkel tree of the block transaction, and the timestamp \cite{nakamoto_bitcoin_2009}. The block body includes verified transactions and a complete Merkel tree composed of these transactions \cite{Szydlo_2004}. The Merkel tree is a binary tree, where the bottom layer corresponds to the content of the leaf node. Each leaf node is the hash value of the corresponding data. Two neighboring leaves unite to perform a hash computation that becomes the content of the upper-level node. Recursive computations form the content of the root node. Blockchain’s non-tampering is ensured by Merkle tree’s particular data structure since any modification in the leaf will be passed to its parent and propagated to the tree's root \cite{pham2020double}. 
(5) \textbf{Application Layer}: Encapsulates various script codes, smart contracts, Decentralized Applications (DApps) and APIs. Scripts are sets of instruction lists attached to transactions. Smart contracts are event-driven, stateful computer programs that run on a shared blockchain data ledger, processing data and managing on-chain smart assets. DApps are decentralized and secure applications that run on a distributed network, utilizing open-source code and storing data and records on the blockchain using cryptographic technologies. APIs are provided for development, allowing DApps and third-party applications to handle smart contracts and retrieve data from blockchain.
\subsection{Types of blockchain networks}
Blockchain networks can be categorized as public, consortium, or private in order of decreasing degrees of openness available for participation by nodes.

\textbf{Public Blockchain}: Referred to permissionless blockchain, enable any node to enter and exit the network \cite{VRANKEN20171}. It is completely decentralized, and each node can participate anonymously without registration, authorization, or authentication \cite{Yaga_Mell_Roby_Scarfone_2018}. Cryptography-related technologies such as digital signatures, hashing \cite{Dods_Smart_Stam_2005}, symmetric/asymmetric keys \cite{7086640}, and ECDSA \cite{Johnson_Menezes_Vanstone_2001} are used to ensure that transactions cannot be tampered with. Economic incentives such as transaction fees and rewards motivate consensus nodes to participate in the consensus process.

\textbf{Private Blockchain}: Known as the permissioned chain, is generally not open to the outside world and is only used by individuals or institutions \cite{Thakur_Kulkarni_2017}. Access to read and write on the private blockchain is governed by rules established by private organizations. Private chains prioritize preventing internal and external security attacks on data and providing users with a secure, tamper-proof, and traceable system. They offer a certain degree of centralized control instead of complete decentralization, sacrificing some of the latter's benefits for better performance compared to public chains.

\textbf{Consortium Blockchain}: A hybrid architecture comprising features from public and private blockchains in which participation is limited to a consortium of participating members \cite{Consortium}. Each node may refer to a single organization or institution in the consortium. The number of nodes  is determined by the size of the pre-selected participants in the blockchain. For example, a financial blockchain is designed for a consortium of 30 financial institutions and allows 30 nodes in this consortium blockchain. The number of nodes required to reach a consensus depends on which algorithm the consortium blockchain uses. The consortium chain accesses the network through the gateways of member institutions and generally provides members’ information authentication, data read and write permission authorization, network transaction monitoring, member management, and other functions. Each member can have permissions assigned by the consortium to access the ledger and validate the generation of blocks. The Hyperledger project is an example of consortium blockchain architecture. Since there are relatively few nodes participating in the consensus process, the consortium blockchain generally does not use the PoW mining mechanism as the consensus algorithm. Consortium blockchain requirements for transaction confirmation time and transaction throughput are, thus, different from those of public blockchains.

\subsection{Consensus Algorithms}
\input{table/T_consortium}
A consensus algorithm ensures the integrity of a distributed ledger by facilitating agreement among nodes on transaction content and order. Without it, data inconsistency can occur, leaving the ledger vulnerable to manipulation. Consensus algorithms can address several blockchain problems, including:

\textbf{The CFT Problem}: CFT consensus algorithms only guarantee a blockchain’s reliability and resiliency to blockchain node failure \cite{Nguyen_Survey}. Also known as non-Byzantine errors, node failures can be caused by failed hardware, crashed processes, broken networks, or software bugs. CFT cannot address scenarios involving malicious activities, referred to as Byzantine errors. When nodes intentionally and maliciously violate consensus principles, e.g., tampering with data, a CFT algorithm cannot guarantee the system reliability. Thus, CFT consensus algorithms are mainly used in closed environments such as enterprise blockchains. Current mainstream CFT consensus algorithms include Paxos and Raft. The latter is a derivative of the former and a simplified consensus algorithm designed to be more suitable for industry implementation.
 
\textbf{The BFT Problem}: Unlike CFT problems that deal with crashes or failures, a Byzantine fault is caused by malicious nodes sending incorrect information to prevent other nodes from reaching consensus. In distributed systems, the Byzantine General's Problem translates into an inability to maintain consistency and correctness under certain conditions. The Byzantine Generals Problem, proposed by Lamport \cite {lamport_byzantine_1982}, is described as follows. Several Byzantine armies are camping outside an enemy city, and a general commands each army. The generals can only communicate with each other by dispatching a messenger who carries messages back and forth \cite{lamport_byzantine_1982}. After assessing the enemy’s situation, they must agree on an identical action plan. However, some traitors among these generals may prevent loyal generals from reaching an agreement. The generals require an algorithm to guarantee that all loyal generals reach a consensus, even if a small number of traitors cheat.
\begin{figure}[htp]
    \centering
    \includegraphics[width=\linewidth]{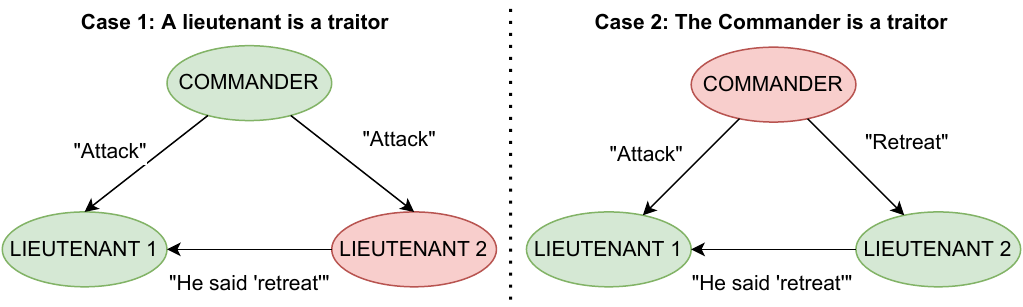}
    \caption{Byzantine General Problem}
    \label{fig:bft}
\end{figure}
Let $v(i)$ represent the information sent by the $i$-th general. Each general draws up a battle plan based on $v(1)$, $v(2)$, $\cdots$, $v(n)$, where $n$ is the number of generals. The problem can be described in terms of how a commanding general issues orders to lieutenants. Therefore, the problem will be transformed into the following \textbf{Byzantine General Problem}: A commander issues an order to his $n-1$ lieutenants such that: (a) \textsf{IC1}. All loyal lieutenants obey the same order. (b) \textsf{IC2}. If the commander is loyal, each loyal lieutenant must obey the order. \textsf{IC1} and \textsf{IC2} are conditions for interactive consistency, which is a configuration that includes the number of generals in a final agreement \cite{lamport_byzantine_1982}.

One case of the Byzantine Generals Problem is shown in Figure \ref{fig:bft}. Here, the Commander and Lieutenant 1 are loyal, and Lieutenant 2 is a traitor. The Commander issues an attack order to all lieutenants. Lieutenant 2 is a traitor, and he/she deceives Lieutenant 1 by sending a tampered message called ``retreat". Since Lieutenant 1 does not know whether the Commander or Lieutenant 2 is a traitor, he/she cannot judge which message includes the correct information and, thus, cannot reach a consensus with the loyal Commander.
In another case, shown in Figure \ref{fig:bft}, the two lieutenants are loyal, and the Commander is a traitor. The Commander issues different orders to the two lieutenants. Lieutenant 2 conscientiously delivered the information of the Commander to Lieutenant 1. Lieutenant 1 cannot judge which information is correct, resulting in loyal lieutenants not reaching a consensus.


If there are $f$ traitors and the total number of generals $n$ is less than $3f + 1$, the Byzantine Generals Problem has no solution. Lamport \cite {lamport_byzantine_1982}  proposed a solution to solve the Byzantine Generals Problem in exponential time $\mathcal{O}(n^f)$ if the adversary mode is $n = 3f + 1$. This original BFT algorithm is computationally expensive to implement, and a practical BFT algorithm is introduced later. 

The \textbf{adversary model} represents a malicious entity that aims to prevent non-malicious entities from achieving their goal \cite{Adversary_2020}. An adversary model imposes a specific limit on the percentage of computing power or property that an adversary can hold, generally represented by $f$ for the number of adversaries and $n$ for the total number of nodes. For example, if a BFT algorithm’s adversary model is $n = 3f + $1, it implies that if the algorithm can tolerate $f$ faulty replicas, the system requires a minimum number of $n = 3f + 1$ replicas.

\subsection{Consensus algorithm classification}
One way of classifying consensus algorithms is by their approach to making final decisions to reach consensus \cite{Nguyen_Survey}. The first category is proof-based consensus algorithms, since a node in this category has to compete with other nodes and prove that it is more qualified than others to commit transactions. PoW \cite{nakamoto_bitcoin_2009}, PoS \cite{pos_paper}, Proof of Authority (PoA) \cite{VeChain_Whitepaper}, Proof of Elapsed Time (PoET) \cite{poet}, and Proof of Space (PoSpace) \cite{cryptoeprint:2013:796} are algorithms of this category. The other category is that of voting-based algorithms since the commitment depends on which committed result wins the majority of votes. Paxos \cite{Paxos_paper}, Raft \cite{Raft_paper}, PBFT \cite{Pbft_paper}, RFBT \cite{Rbft_paper}, RPCA \cite{Rpca_paper}, SCP \cite{SCP_paper}, Tendermint \cite{Tendermint}, and HotStuff \cite{Hotstuff_paper} belong to this category. The first group of consensus algorithms is proof-based, while the second group is voting-based. Another way of classifying consensus algorithms is by the design principle of fault tolerance. Nodes can suffer from non-Byzantine errors (also known as crash faults), which is exemplified by situations where the node fails to respond. 
\begin{figure*}[t!]
    \centering
    \includegraphics[width=.65\textwidth]{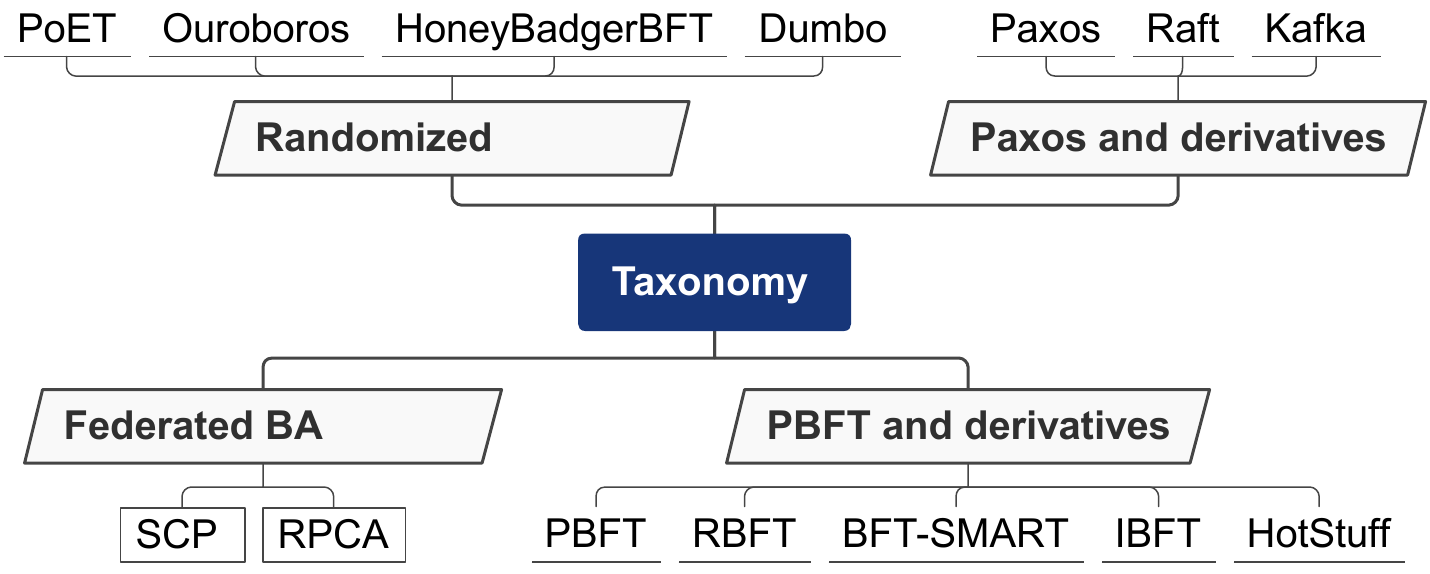}
    \caption{Proposed Taxonomy for Assessing Consensus}
    \label{fig:proposedTaxonomy}
\end{figure*}
Alternatively, nodes can forge or tamper with the information and respond maliciously, causing Byzantine Fault. Thus, consensus algorithms may be classified as being designed for Crash Fault Tolerance (CFT) or Byzantine Fault Tolerance (BFT). This classification method only focuses on the original design principle. Most BFT-based consensus algorithms can tolerate either crash fault or Byzantine fault. Since the design principle of algorithms in the previous proof-based family is different from fault tolerance, those proof-based families will be excluded from this classification. Paxos \cite{Paxos_paper}, Raft \cite{Raft_paper}, and Zab \cite{Junqueira_Reed_Serafini_2011} belong to the category of CFT-based consensus algorithms. A number of variants of PBFT \cite{Pbft_paper} algorithms, such as RBFT \cite{Rbft_paper}, SBFT \cite{Gueta_Abraham_Grossman_Malkhi_Pinkas_Reiter_Seredinschi_Tamir_Tomescu_2019}, BFT-SMART \cite{Bftsmart_paper}, DBFT \cite{dbft_paper}, and HotStuff \cite{Hotstuff_paper} form a collection in the category of BFT-based consensus algorithms. Another group of consensus algorithms, forming a collection in the same category, uses Byzantine Federated Agreement (BFA) \cite{SCP_paper} for voting, such as RPCA \cite{Rpca_paper} and SCP \cite{SCP_paper}.

In the next section, we will propose a more comprehensive taxonomy, aiming to better capture the nuances and complexities of the evolving landscape of consensus algorithms.

%% file: table/T_consortium.tex
\begin{table*}
\caption{Mainstream Platforms and Consensus Algorithms}
\label{tab:PlatformsConsensus}
    \begin{tabular}{p{0.15\textwidth}p{0.18\textwidth}p{0.17\textwidth}p{0.11\textwidth}p{0.05\textwidth}p{0.09\textwidth}p{0.08\textwidth}}
    \toprule
    \textbf{Blockchain Platforms} & \textbf{Consensus Algorithm} & \textbf{Use Cases Areas} & \textbf{Smart Contract} & \ \textbf{DApps} & \textbf{Open Source} & \textbf{Open APIs}\\
    \midrule
    Antchain & HoneybadgerBFT \cite{miller2016honey} & Multi-Purpose & x & \checkmark  & x & \checkmark \\
    Cardano & Ouroboros \cite{ou_paper} & Cryptocurrency & \checkmark & \checkmark  & \checkmark & \checkmark \\
    Enterprise Ethereum	& Customized & Multi-Purpose & \checkmark & \checkmark & \checkmark & \checkmark \\
    FISCO BCOS 	& Raft \cite{Raft_paper}, PBFT \cite{Pbft_paper} & Multi-Purpose & \checkmark & \checkmark & \checkmark & \checkmark \\
    Google Chubby & Paxos \cite{Paxos_paper} & Distributed System & x & x & \checkmark & \checkmark \\
    Hedera & Hashgraph \cite{Hashgraph_paper} & Cryptocurrency &\checkmark &\checkmark & x & \checkmark \\
    Hyperledger Besu & QBFT, IBFT \cite{ibft2_paper}  & Multi-Purpose & \checkmark & \checkmark& \checkmark & \checkmark \\
    Hyperledger Burrow & Tendermint \cite{Tendermint_paper} & Multi-Purpose & \checkmark & \checkmark & \checkmark & \checkmark \\
    Hyperledger Fabric & PFBT, Raft \cite{Raft_paper}, Kafka \cite{kafka_paper} & Supply chain & \checkmark &\checkmark& \checkmark & \checkmark \\
    Hyperledger Firefly	& IBFT \cite{ibft2_paper} & Multi-Purpose & \checkmark & \checkmark & \checkmark & \checkmark\\
    Hyperledger Indy & RBFT \cite{Rbft_paper} & Identity Management & x & \checkmark & \checkmark & \checkmark \\
    Hyperledger Iroha & YAC \cite{yac_paper} & Multi-Purpose & x & \checkmark  & \checkmark & \checkmark \\
    Hyperledger Sawtooth & PoET \cite{Poet_paper} & Multi-Purpose & \checkmark & \checkmark & \checkmark & \checkmark \\
    IPFS private & Raft \cite{Raft_paper} & Data Storage & \checkmark & \checkmark & \checkmark & \checkmark \\
    Libra & LibraBFT \cite{libra_association_members_libra_2020} & Cryptocurrency	& \checkmark & \checkmark & x & \checkmark \\
    MOBI & MOBI Consensus & IoT & \checkmark & \checkmark & x & \checkmark \\
    Neo	& dBFT \cite{dbft_paper} & Cryptocurrency & \checkmark & \checkmark  & x & \checkmark \\
    Quorum	& IBFT, Kafka \cite{kafka_paper} & Financial  & \checkmark & \checkmark & x & x \\
    R3 Corda & Customized & Multi-Purpose & \checkmark & \checkmark & \checkmark & \checkmark \\
    Ripple	& RPCA \cite{Rpca_paper} & Cryptocurrency, Financial & x & x & x & \checkmark \\
    Stellar	& SCP \cite{SCP_paper} & Cryptocurrency, Financial & x & \checkmark & x & \checkmark \\
    Symbiont & BFT-SMART \cite{Bftsmart_paper}  &  Fintech & x & x & x & x \\
    Tendermint	& Tendermint \cite{Tendermint_paper} & Multi-Purpose  & \checkmark & \checkmark & x & \checkmark \\
    TradeLens & Raft \cite{Raft_paper} & Supplychain & \checkmark & \checkmark & x & \checkmark \\
    Trusted IoT Alliance & Customized & IoT & \checkmark & \checkmark & x & \checkmark \\	
    VeChain	& Proof of Authority \cite{de2018pbft} & Supplychain & x & \checkmark & x & \checkmark \\
    Zookeeper & Paxos \cite{Paxos_paper}, Kafka \cite{kafka_paper} & Distributed System & x & x & \checkmark & x \\
    \bottomrule
    \end{tabular}
\end{table*}

%% file: 3_taxonomy.tex

\section{Taxonomy of Consensus Mechanisms} \label{sec:taxonomy}
Mainstream consortium blockchains and distributed systems \cite{publicblockchainsurvey, Thakur_Kulkarni_2017, casino2019systematic, zou2020focus, Consortium, Nguyen_Survey, Salimitari_Chatterjee_2019,Xiao_Zhang_Lou_Hou_2019, Cachin2017BlockchainCP,Ferdous2020BlockchainCA, bellaj2022sok} considered in this paper are shown in Table \ref{tab:PlatformsConsensus} along with their consensus algorithms, use cases areas, and availabilities for smart contract, DApps, open source, and open APIs. From Table \ref{tab:PlatformsConsensus}, a representative list of consensus algorithms used in mainstream consortium blockchains are selected based on their methodologies of achieving consensus. Figure \ref{fig:proposedTaxonomy} shows our proposed taxonomy of consensus algorithms.



\subsection{Paxos and derivatives}
Paxos \cite{Paxos_paper} is a classical and widely used consensus algorithm in distributed systems. Due to its efficiency and usability, its derivatives are adopted in many blockchain platforms.
\subsubsection{Paxos}
\input{algorithms/paxos}
\subsubsection{Raft}
\input{algorithms/raft}
\subsection{PBFT and derivatives}
PBFT is the practical BFT algorithm for enterprise-level distributed systems. Numerous variants are proposed to improve its ability to solve more challenges in blockchain.
\subsubsection{Practical Byzantine Fault Tolerance (PBFT)}
\input{algorithms/pbft}
\subsubsection{Redundant Byzantine Fault Tolerance (RBFT)}

\input{algorithms/rbft}

\subsubsection{BFT-SMART}
\input{algorithms/bftsmart}
\subsubsection{HotStuff}
\input{algorithms/hotstuff1}

\subsubsection{LibraBFT}
\input{algorithms/librabft}
\subsection{Federated Byzantine Agreement}
The algorithms in this category differ from PBFT variants in that a group of nodes can choose one or more nodes as representatives.
\subsubsection{RPCA}
\input{algorithms/rpca}

\subsubsection{Stellar Consensus Protocol (SCP)}
\input{algorithms/scp}

\subsection{Randomized consensus mechanisms}
The algorithms in this category leverage aspects of randomized mythologies to choose nodes that have decisive effects on reaching consensus in the network. 
\subsubsection{PoET} 
Proof of Elapsed Time \cite{Poet_paper} is an efficient consensus protocol utilizing a Trusted Execution Environment (TEEs), i.e., Intel SGX-enabled CPUs \cite{costan2016intel}, implemented in Hyperledger Sawtooth Project \cite{poet}. Although PoET looks like a competitive consensus algorithm (i.e., PoW) by name, it essentially selects the validator node by a randomized waiting time, in contrast to PoW, where all the nodes compete to solve a cryptographic puzzle and mine the next block. As a result, significantly fewer computational resources are required. In this mechanism, each node waits for a particular random time $\mathcal{T}$ which satisfies a predefined probability distribution function $\mathcal{F}$. The first node to finish the waiting time becomes the primary node. The other nodes verify that the primary node has indeed waited for the particular random time $\mathcal{T}$ in three steps: (a) Proof of waiting time is generated with the assistance of SGX while generating the block on the primary node. (b) The primary node broadcasts the generated proof along with the block to other nodes. The other nodes will produce block and waiting time verification. (c) A probabilistic statistical verification is applied to check whether the node's waiting time meets the predefined probability distribution $\mathcal{F}$. The PoET consensus algorithm relies on the hardware for trust so that the blockchain system does not need to spend significant computing power and achieves fairness. 

\subsubsection{Ouroboros} Based on the PoS \cite{pos_paper}, Ouroboros randomly chooses a stakeholder and authorizes the selected stakeholder permission to generate the block in a certain period \cite{ou_paper}. Ouroboros divides time into epochs, with each epoch containing multiple slots, and at most, one block will be generated per slot. For each slot, a leader is elected, generating a block and passing it to the next slot leader. The process of the slot leader election is randomized, and it has been proven as an unpredictable process. It isdenoted as $\textsf{F}(S,p,sl_j)\to S_i$. Where $S$ is the set of candidate stakeholders, $p$ is the random seed, $sl_j$ denotes the $j$-th slot in the epoch, and $S_i$ the selected node. The randomized algorithm selects the slot leader from the candidate set $S$ for each slot. During the epoch cycle, $S$ and $p$ remain unchanged until the of the cycle. If a new epoch starts, then a new $p$ is generated along with an updated $S$. Each node executes the random function \textsf{F} to check who is the slot leader of the current slot based on the seed $p$ of the current epoch. If the slot leader is itself, then the node will combine the transactions and create a new block; otherwise, it waits for the slot leader to create the block, broadcasts it, and verifies the received block. If a node has not received a broadcast for a long time (beyond the time of one slot), it considers that no block is generated for current slot. This process is repeated in current epoch until all slots are finished.

In Ouroboros, the random seed $p$ for each epoch is generated by a secure multi-party computation protocol (MPC), which uses a publicly verifiable secret-sharing algorithm \cite{stadler1996publicly} to guarantee the generation of bias-resistant random numbers in the presence of adversaries. Similar to the voting algorithms, the randomly generated leader node reduces resource consumption. However, unlike the voting algorithms, in which the generation of leaders requires network communications between nodes, the randomized algorithm uses a random algorithm to generate leaders without communication, which shortens the running time for reaching consensus, improves efficiency, and enhances scalability.
 
\subsubsection{HoneyBadgerBFT (HB-BFT)}
HB-BFT \cite{miller2016honey} achieves distributed consensus consistency in asynchronous networks without requiring any time assumptions, and solves the transaction censorship problem through the use of threshold encryption. In each consensus round, the protocol operates on data of size $B$, assuming $n$ nodes in the network. The primary process of HB-BFT is as follows:
(a) Each node collects transaction data in its own transaction data buffer. Before each round, the node removes the first $B/n$ transactions from the buffer. HB-BFT then divides a block into $n$ copies, each containing different transactions. After that, these copies will be sent to replica nodes. Then, replica nodes will exchange the remaining $n-1$ copies to fully use the bandwidth between replica nodes and ease the broadcast pressure of the primary node. 
(b) Each replica node randomly selects transactions to verify, sign, and broadcast with threshold encryption to improve transaction throughput. As HB-BFT is an asynchronous consensus protocol, the transactions received by each replica node are asynchronous and random, and their arrival order can vary. After receiving transaction information, each node stores it in a local cache pool and randomly selects several transactions to process. The node then verifies, signs, and broadcasts the selected transactions with threshold encryption using Reliable Broadcast (RBC). The ciphertext data is then generated and broadcasted by the replica. Eventually, each node combines the different subsets of transactions received to form the complete set of transactions contained in the block of that view.
(c) The final confirmation of required transactions is achieved through Asynchronous Binary Byzantine agreement (ABA). The primary node forms a transaction set ciphertext from encrypted data received from nodes and initiates ABA consensus. A consensus round determines the final confirmed binary value, which determines which transactions will be approved. Nodes can confirm the Asynchronous Common Subset (ACS) subcomponent after eliminating invalid and duplicate transactions based on the received value $V$.
(d) Transaction data threshold decryption. When the ABA consensus is completed, i.e., the transaction set's ciphertext data is confirmed, the nodes run the threshold decryption algorithm. As long as at least $f + 1$ nodes complete the decryption, the plaintext data in the set can be restored, and the transaction is confirmed. There are two advantages to this strategy: first, it prevents adversaries from understanding the transaction selection strategy to interfere or attack; and second, although the order of transactions received by each node may not be consistent, most of them may be in the same order under similar network conditions, and random selection instead of all sequential selection can avoid a large number of duplicates.

\begin{figure*}[!t]
    \centering
    \includegraphics[width=.8\textwidth]{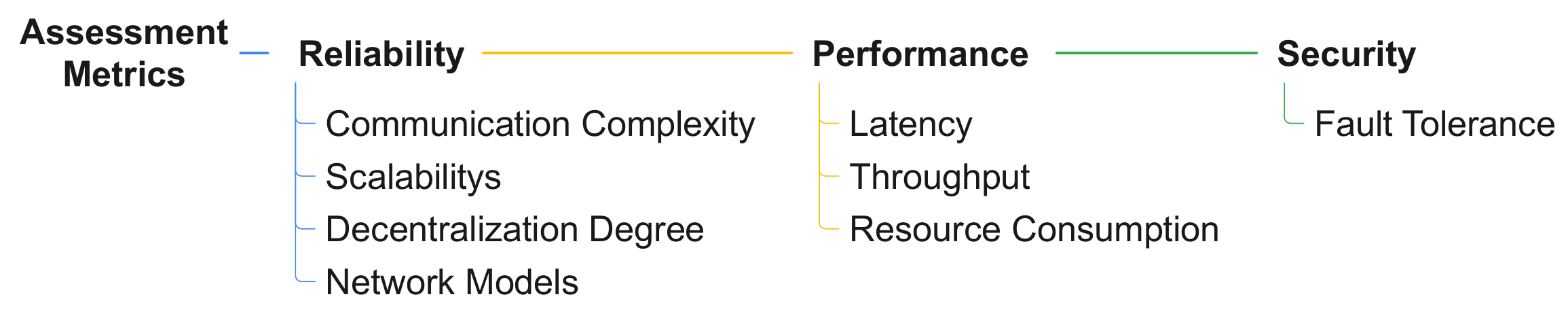}
    \caption{Assessment Metrics for Proposed Taxonomy}
    \label{fig:compMethologies}
\end{figure*}
\subsubsection{Dumbo} Proposed as the first practical asynchronous BFT protocol \cite{Dumbo_paper} based on HB-BFT and the motivation to address performance bottleneck of HB-BFT due to the number of ABA instances. Because each node has to run $N$ instances of ABA in each round, and each instance has to verify $\mathcal{O}(n^2)$  threshold signatures, HB-BFT consumes a lot of computing resources. To improve the efficiency of asynchronous consensus by reducing the number of ABA instances in each round, Dumbo uses two atomic broadcast protocols: (1) Dumbo1 which reduces the number of instances running ABA from $N$ to $k$ by choosing a committee of $k$ members. By leveraging an added coin-tossing protocol, Dumbo1 can randomly elect a committee of $k$ members, with a high probability that at least one of the $k$ members is honest. (2) Dumbo2, which reduces number of ABA to a constant running time by using the Multi-value Validated Byzantine Agreement (MVBA) \cite{miller2016honey}. To apply MVBA, a provable reliable broadcast (PRBC) is implemented by levering threshold signature on the RBC index. As a result, only three consecutive ABA instances need to be performed.

\subsection{Assessment Methodology}
Selected consensus algorithms are compared along the following three dimensions that form the backbone of scalable solutions in blockchain technology: (1) Reliability is measured by communication complexity, scalability, and decentralization. (2) Performance is measured by latency, throughput, and resource consumption. (3) Security is measured by fault tolerance. Figure \ref{fig:compMethologies} shows comparison methodologies for proposed taxonomy. The Assessment for selected algorithms will be presented in section \ref{sec:reliability}, \ref{sec:performance}, and \ref{sec:security}.

%% file: algorithms/paxos.tex
A fault-tolerant consensus algorithm that relies on message passing in a distributed system \cite{Paxos_paper}. It divides nodes into three roles: \textbf{proposer}, \textbf{acceptor}, and \textbf{learner}. Each node can have multiple roles simultaneously. 
A proposer is responsible for presenting a proposal and awaiting acceptors' responses. An acceptor votes on the proposal. A learner is informed of the proposal’s result but does not participate in voting. Paxos ensures the proposal's uniqueness with a proposal number and content with a value. When more than half of the acceptors approve a proposal, it is considered \textsf{Chosen}. 
Paxos ensures safety and liveness, ensuring the proposed value is chosen and the proposal is completed within a limited time.

\begin{figure}[htp]
    \centering
    \includegraphics[width=\linewidth]{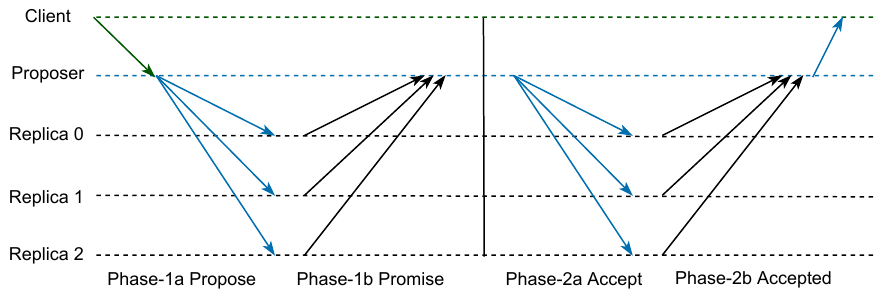}
    \caption{Paxos two phases.}
    \label{fig:paxos}
\end{figure}

Paxos execution is divided into two phases, as shown in figure \ref{fig:paxos}. 
In the \textbf{Prepare} phase, the proposer sends a \textsf{Prepare} request with a proposal number to more than half of the acceptors in the network to determine whether a majority is prepared to accept the proposal. After receiving the proposal, the acceptor stores the largest proposal number it has received and returns a \textsf{Promise} message to the proposer if the proposal number is greater than the saved maximum proposal number. The acceptor promises not to accept any proposal with a number less than the proposal number that has already been received.
In the \textbf{Accept} phase, the proposer broadcasts an \textsf{Accept} request with the proposal if it receives more than half of the responses as \textsf{Promise} messages. The request contains a proposal number and the value that the node would like to propose. If the response does not contain any proposal, the proposer determines the value. If the response message contains a proposal, the value is replaced by the value in the response with the largest proposal number. 
After an acceptor receives the \textsf{Accept} request, it accepts the proposal and updates the accepted maximum proposal if the proposal number is not less than the maximum proposal number promised by the acceptor. 
If a majority of acceptors accept the proposal, the value is \textsf{Chosen}, indicating consensus has been reached.


%% file: algorithms/raft.tex
Motivated by Paxos, Raft is designed for ease of understandability and implementability for industry applications \cite{Raft_paper}. Its core premise is that servers start from the same initial state and execute a series of command operations in the same order. Its goal is to achieve a consistent state, and therefore, it uses the log method for synchronization, a consistent algorithm for managing replicated logs. 
Raft divides nodes into three mutually convertible roles: \textbf{leader}, \textbf{follower}, and \textbf{candidate}, with only one leader in a cluster of at least five nodes.
The leader handles client requests, replication logs, and communication with followers.
Initially, all nodes are followers, which passively respond to Remote Procedure Call (\textsf{RPC}) requests from the leader. Followers do not communicate with each other but instead respond to log replication and election requests from the leader and candidate nodes, respectively. If a follower receives a request from a client, it forwards it to the leader.
A candidate initiates an election vote when the leader becomes inactive, and one or more nodes may switch from follower to candidate. Once a candidate wins the election, it becomes the new leader and can revert to a candidate if a new leader is elected but then fails.

Raft runs in two phases. The first phase is leader election, where a leader sends \textsf{heartbeat} messages to followers to maintain its authority. If a follower does not receive the heartbeat within an \textsf{Election timeout} period, it switches to a candidate role and starts an election process, signaling that the leader has failed \cite{Raft_paper}. Then, it increases its term, sends \textsf{RequestVote RPC} to other servers, and waits for results. If a candidate wins the election, it becomes the leader. If another server wins the election, the candidate becomes a follower. If no one wins, the election is reinitiated. The second phase is log replication, where the leader accepts client requests, updates log, and sends a heartbeat to followers to synchronize leader's log.

%% file: algorithms/pbft.tex
It is a consensus algorithm based on state machine replication \cite{Pbft_paper} where services are replicated on different nodes in a distributed system. Each copy of the state machine saves the state of the service and the operations. PBFT reaches consensus through three phrases: \textbf{Pre-prepare}, \textbf{Prepare}, and \textbf{Commit}. In PBFT, there is one primary node out of $n$ nodes, and others are backup. If the primary node fails, the backups nodes initiate a view-change to select a new primary node.


The process of reaching consensus in PBFT is as follows: 
(a) \textbf{Propose}: The client uploads the request message $m$ to the primary node and replicas. 
(b) \textbf{Pre-prepare}: The primary node receives the request message $m$ and generates a \textsf{Pre-prepare} message $\langle \textsf{Pre-prepare}, \textsf{H}(m), s, v\rangle$, where $\textsf{H}(m)$ is a one-way hash function, $s$ is the message sequence number, and $v$ represents the current view. The primary node signs the message with its private key and sends it to replicas.
(c) \textbf{Prepare}: Upon receiving the \textsf{Pre-prepare} message, replicas verify the message and create a \textsf{Prepare} message $\langle \textsf{Prepare}, \textsf{H}(m), s, v\rangle$ and broadcast it to the network. If a replica node receives $2f + 1$ valid \textsf{Prepare} messages, it generates a prepared certificate.
(d) \textbf{Commit}: If a replica has a prepared certificate, it broadcasts a \textsf{Commit} message $\langle \textsf{Commit}, s, v \rangle$ and stores the message $m$ in the local log. If a replica receives $2f + 1$ valid \textsf{Commit} messages, it generates a committed certificate, indicating the message has been successfully committed.
(e) \textbf{Reply}: Once a node receives $2f+1$ valid \textsf{Commit} messages, it sends the committed certificate and the message $m$ to the client.
PBFT includes a view-change protocol in case of primary node failure, where a replica triggers a view change to elect a new primary node if it has not received a response from the current primary after a timeout, ensuring liveness.
\begin{figure}[htp]
    \centering
    \includegraphics[width=\linewidth]{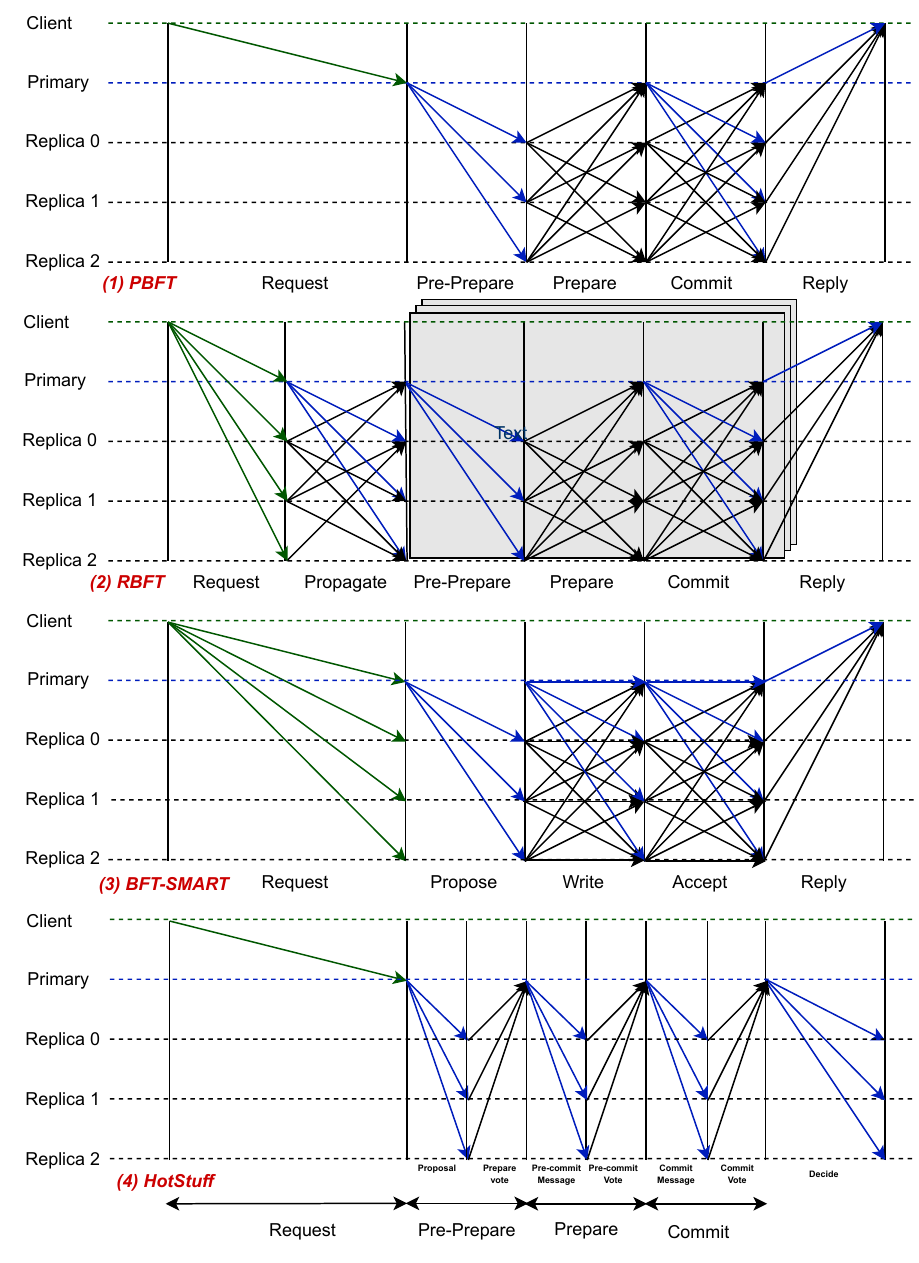}
    \caption{Processes of PBFT algorithm and its derivatives}
    \label{fig:bftstyle}
\end{figure}

%% file: algorithms/rbft.tex
A variant of PBFT that improves robustness by using a multi-core architecture \cite{Rbft_paper}. In RBFT, each node runs $f + 1$ PBFT protocol instances in parallel, with only one instance serving as the master and the rest as backup. Each instance has its own $n$ replicas, and in $f + 1$ instances, each node can have at most one primary. 

RBFT uses a communication process similar to PBFT in the consensus protocol phase, but adds a propagate phase before the \textbf{Pre-prepare} phase to ensure all correct nodes eventually send the request to the next phase, as shown in Figure \ref{fig:bftstyle}. $f+1$ PBFT instances must receive the same client request for correctness, which is achieved by forwarding the message to each other. Once a node receives $2f+1$ requests from a client, it sends the request to $f+1$ instances and moves to the next phase, following the 3-phase process similar to PBFT \cite{Pbft_paper}, and is represented in steps 3, 4, and 5 in Figure \ref{fig:bftstyle}. The algorithm is performed by the $f+1$ instances during the consensus protocol, and the result is returned to the client through MAC authentication messages. When the client receives $f+1$ valid and consistent replies, it accepts these replies as the result.

RBFT improves upon PBFT by implementing a monitoring mechanism and a protocol instance change mechanism to promote robustness \cite{Rbft_paper}. Each node runs a monitoring program to monitor the throughput of all $f+1$ instances, and if $2f+1$ nodes find that the performance difference between the master and the best backup instance reaches a certain threshold, then the primary of the master instance is considered as a malicious node. A new primary is then selected, or the primary in the backup instance with the best performance is chosen, and it is elevated to master. To update all instances' primaries, each node maintains a counter to record the change information of each instance. If a node needs to change the primary, it sends an \textsf{Instance-change} message with a MAC authenticator to all nodes. After receiving the incoming \textsf{Instance-change} message, each node verifies the MAC and compares it with its counter. If the counter is larger, it discards the message. Otherwise, the node checks whether it also needs to send the \textsf{Instance-change} message by comparing the performance of the master and backup. If $2f+1$ valid \textsf{Instance-change} messages are received, the counter is incremented by one, and this starts the view-change process to update all instances' primaries, including the master's.

%% file: algorithms/bftsmart.tex
A state machine replication library written in the Java \cite{Bftsmart_paper}. It supports state transfer services to repair a failed node, add/remove replicas through a Trusted Third Party client, and access other nodes to obtain the latest replica status. The system ensures stable error recovery from $f$ faulty nodes by storing each node's operation logs on other disks. 
BFT-SMART also divides nodes into leaders and backups and employs a reconfiguration protocol \cite{Alchieri_Dotti_Mendizabal_Pedone_2017} similar to the view-change protocol in PBFT to handle failure.

The consensus process in Mod-SMaRt \cite{Sousa_Bessani_2012} is based on a leader-driven algorithm \cite{Cachin} with three phases: \textbf{Propose}, \textbf{Write}, and \textbf{Accept}, as shown in Figure \ref{fig:bftstyle}. 
A leader is elected from the network. When a client initiates a request, it sends a \textsf{Request} message containing the client serial number, digital signature, and operation request content to all nodes and waits for a response. In the normal phase, the leader verifies the correctness of the received \textsf{Request} message. 
After verification, the leader accepts the message, assigns a serial number, and sends the \textsf{Request} message to the replicas. As long as a replica accepts the message and forwards it, other nodes will also receive and send the \textsf{Write} message to all nodes, including itself. When a node receives $2f$ \textsf{Write} messages, the node broadcasts an \textsf{Access} message to all nodes, including itself. When a node receives $2f+1$ \textsf{Access} messages, the request is executed. The protocol stores content of the series of request operations and the encrypted certificate in each node's log and replies \textsf{Access} to client.


In BFT-SMaRt, if a node experiences two timeouts, it enters the synchronization phase and the reconfiguration protocol re-elects the leader node. 
During first timeout, the consensus and reconfiguration processes can execute simultaneously. The \textsf{Request} message is automatically forwarded to all nodes after the first timeout, and a \textsf{Stop} message is sent to other nodes after the second timeout. Once a node receives more than $f$ \textsf{Stop} messages, it starts the next reconfiguration phase immediately. 
After the leader election, all nodes send a \textsf{Stopdata} message to the new leader. If the leader accepts at least $n-f$ valid \textsf{Stopdata} messages, it sends a \textsf{Sync} message to all nodes. Replicas start to synchronize if the leader is valid.

%% file: algorithms/hotstuff1.tex
A partially synchronized network \cite{Chan_Pass_Shi} with an adversary model of $n = 3f + 1$. It uses a parallel pipeline to process a proposal, which is equivalent to combining the preparation and commitment phases of PBFT. The original proposal includes two implementations of HotStuff, Basic HotStuff, and Chained HotStuff. The Basic HotStuff protocol forms the core of HotStuff, which switches between a series of views. The views switch according to a monotonically increasing number sequence. A unique consensus leader exists within each view. Each replica node maintains a tree structure of pending commands in its memory. Uncommitted branches compete, and only one branch in a round will be agreed upon by the nodes. In the HotStuff protocol, branches are committed as the view number grows. Voting in HotStuff uses the cryptographic term \textsf{Quorum Certificate} (\textsf{QC}), where each view is associated with a \textsf{QC} that indicates whether enough replicas have approved the view. If a branch is confirmed by a replica, it signs the branch, creates a partial certificate \cite{Chan_Pass_Shi}, and sends to the leader. The leader collects $n - f$ partial certificates, which can be combined into a \textsf{QC}. A view with a \textsf{QC} means that it received the majority votes of the replicas. The leader collects signatures from $n-f$ replicas by using threshold signatures \cite{cachin_blockchain_2017,Gueta_Abraham_Grossman_Malkhi_Pinkas_Reiter_Seredinschi_Tamir_Tomescu_2019}. The process of collecting signatures consists of three phases: \textbf{Prepare}, \textbf{Pre-prepare}, and \textbf{Commit}. Moreover, the entire algorithm consists other two phases: \textbf{Decide} and \textbf{Finally}, as shown in Figure \ref{fig:bftstyle}. 

(1) \textbf{Prepare}. The leader denoted by the current highest view designated as \textsf{highQC}, initiates a proposal for \textsf{highQC}, encapsulates it into a \textsf{Prepare} message with content $\langle \textsf{Prepare}, \textsf{CurProposal}, \textsf{HighQC}\rangle$, and broadcasts it to all replicas. Replicas will decide whether to accept the proposal or not, and then return a vote with partial signature to the leader if the proposal is accepted. (2) \textbf{Pre-commit}. When the leader receives votes from $n-f$ replicas for the current proposal, it combines them into \textsf{PrepareQC}, encapsulates \textsf{PrepareQC} into a \textsf{Pre-commit} message, and broadcasts it to all replicas. The replica votes after receiving the above proposal message and returns the vote to the leader. (3) \textbf{Commit}. When the leader receives the \textsf{Pre-commit} votes from $n-f$ replicas, it merges them into \textsf{PrecommitQC}, encapsulates a \textsf{PrecommitQC} into a \textsf{Commit} message, and broadcasts them to all replicas. The replica votes after receiving the proposal message and returns the \textsf{Commit} vote to the leader. To ensure the safety of the proposal, the replica is locked by setting its \textsf{LockedQC} to \textsf{PrecommitQC}. (4) \textbf{Decide}. When the leader receives the \textsf{Commit} votes from $n-f$ replicas, it merges them into one \textsf{CommitQC} and then uses the \textsf{Decide} message to broadcast it to all replicas. After receiving this message, the replica confirms and submits the proposal in the \textsf{CommitQC}, executes the command, and returns it to the client. After this, the replica increases the \textsf{ViewNumber} and starts the next view. (5) \textbf{Finally}. If the system moves to the next view, each copy sends a message to the next view's leader with the message $\langle\textsf{New-view}, \textsf{PrepareQC}\rangle$.

The processes in each phase of Basic HotStuff are very similar to each other, as shown in Figure \ref{fig:bftstyle}. A modified version of HotStuff, called Chained HotStuff, was proposed \cite{Hotstuff_paper} to optimize and simplify Basic HotStuff. In the Chained HotStuff protocol, the replicas' votes in the \textbf{Prepare} phase are collected by the leader, and stored in the state variable \textsf{GenericQC}. Then, \textsf{GenericQC} is forwarded to the leader of the next view, essentially delegating the next phase's (the \textbf{Pre-commit} phase) responsibilities to the next view's leader. Thus, instead of starting its new \textbf{Prepare} phase alone, the next view's leader actually executes the \textbf{Pre-commit} phase simultaneously. Specifically, the \textbf{Prepare} phase of view $v+1$ also acts as the \textbf{Pre-commit} phase of view $v$. The \textbf{Prepare} phase of view $v+2$ acts as both the \textbf{Pre-commit} phase of view $v+1$ and the \textbf{Commit} phase of view $v$.
\begin{figure}[htp]
    \centering
    \includegraphics[width=8.8cm]{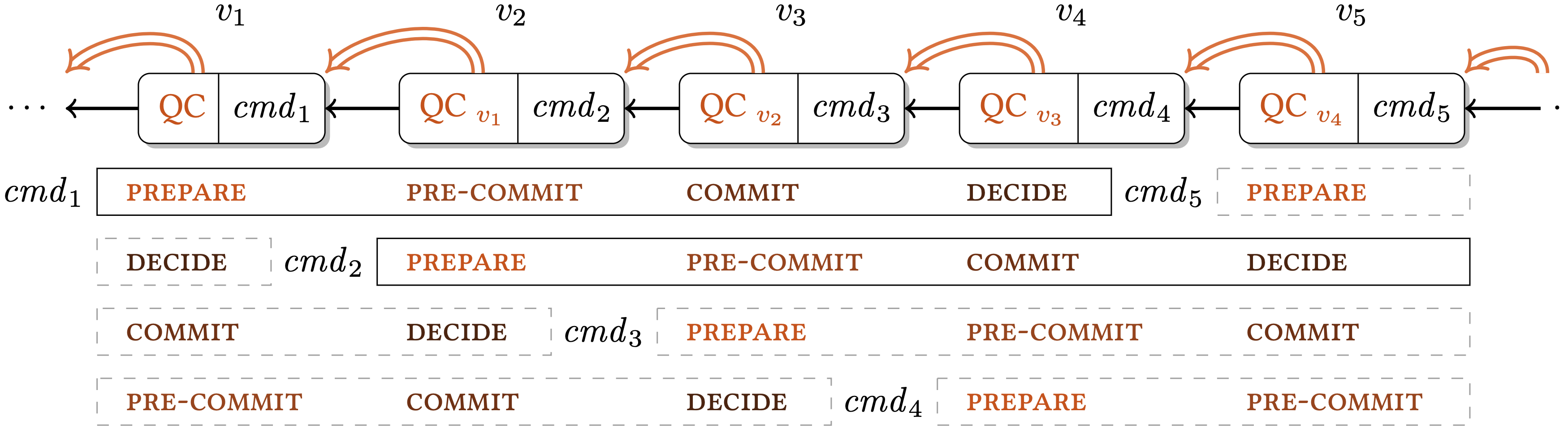}
    \caption{Chained HotStuff is a pipelined Basic HotStuff where a \textsf{QC} can serve in different phases simultaneously. \cite{Hotstuff_paper}} 
    \label{fig:ChaindHotStuff1}
\end{figure}

\begin{figure}[htp]
    \centering
    \includegraphics[width=8.8cm]{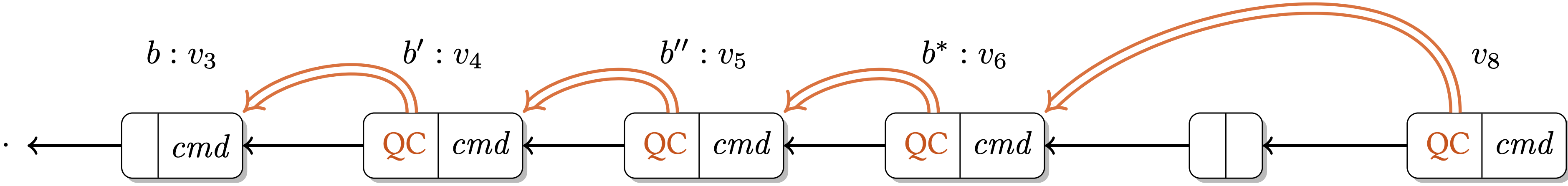}
    \caption{The nodes at views $v_4$, $v_5$, $v_6$ form a Three-Chain. The node at view $v_8$ does not make a valid One-Chain
in Chained HotStuff. \cite{Hotstuff_paper}} 
    \label{fig:ChaindHotStuff2}
\end{figure}

Figure \ref{fig:ChaindHotStuff1} shows that a node can be in different views simultaneously. Through a chained structure, a proposal can reach consensus after three blocks, resembling a Three-Chain shown in figure \ref{fig:ChaindHotStuff2}. An internal state converter enables automatic switching of proposals through \textsf{GenericQC}. The chained mechanism in Chained HotStuff reduces cost of communication messages and allows pipelining of processing. In the implementation of Chained HotStuff, if a leader fails in obtaining enough \textsf{QC}, then it may appear that the view numbers of a node are not consecutive. This is solved by adding dummy nodes, as shown in Figure \ref{fig:ChaindHotStuff2}, where a dummy node is added to force $v_6$, itself, and $v_8$ to form a Three-Chain.

HotStuff achieves $\mathcal{O}(n)$ message authentication complexity by improving the distributed consistency algorithm's efficiency using threshold signatures, parallel pipeline processing, and linear view changing. Compared to PBFT, HotStuff can reach consensus pipelining without a complex view-change mechanism and improves consensus efficiency.

%% file: algorithms/librabft.tex

As a variant of HotStuff, LibraBFT introduces several changes to meet various business requirements. One of the changes is the introduction of epochs, which allows for consensus node replacement and support for incentive and penalty mechanisms \cite{baudet_state_2019}. With the economic incentives and penalties, nodes are encouraged to participate in the voting process and penalized if they violate voting constraints or submit conflicting proposals.
Another change is to address the problem of unknown upper bounds on message latency in HotStuff, which only requires partial synchronization \cite{dwork1988consensus}. The view-change in HotStuff is not time-bound and relies on the status of the last view, which can result in variable confirmation latencies. To address this, LibraBFT uses pacemaker \cite{baudet_state_2019} to ensure that confirmation latency is lower than an upper bound.

%% file: algorithms/rpca.tex
Uses pre-configured validators to vote on transactions for consensus \cite{Rpca_paper}. After several rounds of voting, if a transaction receives a threshold (usually 80\%) of votes, it will be recorded in the ledger. Nodes maintain a subset of validators as a list called Unique Node List (UNL). Non-validators, known as tracking servers, forward transaction information and respond to client requests but don't participate in consensus. A validator and tracking server can switch roles, and inactive validators are removed from the UNL.

\begin{figure}[htp]
    \centering
    \includegraphics[width=\linewidth]{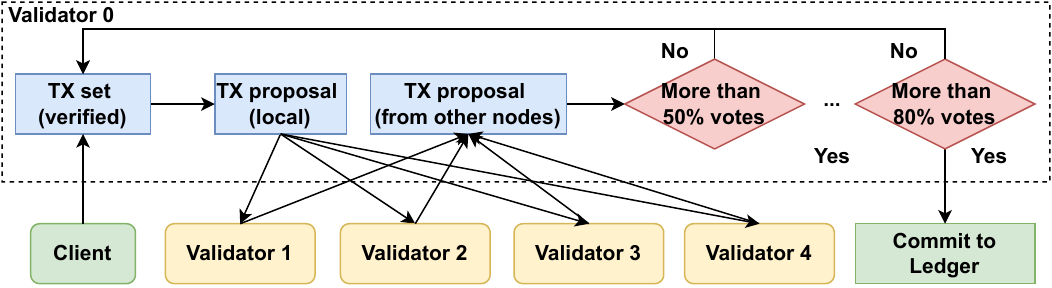}
    \caption{Ripple's RPCA Consensus Algorithm}
    \label{fig:RPCA}
\end{figure}

The process of RPCA is shown in Figure \ref{fig:RPCA}. 
The client initiates a transaction and broadcasts it to the network. Validators receive the transaction data, store it locally, and verify it. Invalid transactions are discarded, while valid transactions are integrated into the candidate set of transactions. Validators periodically send their candidate sets as proposals to other nodes. 
Once a validator receives a proposal, it checks whether the sender is on the UNL. If not, the proposal is discarded. Otherwise, the validator stores the proposal locally and compares it with the candidate set. The transaction obtains one vote if it is the same as in the candidate set.
If the transaction fails to reach 50\% of the votes within a certain period \cite{chase_analysis_2018}, it returns to the candidate set and waits for the next consensus process. If it reaches a threshold denoted by 50\% of votes, it enters the next round and is re-sent as a proposal to other nodes, with the threshold raised.
As the number of rounds increases, the threshold continues to increase until the transaction reaches 80\% or more of the votes, at which point the validator writes it into the ledger. 

%% file: algorithms/scp.tex
A distributed consensus algorithm designed around state machine replication and does not require miners but a distributed server network to run the protocol \cite{SCP_paper}. 
SCP is based on the Federated Byzantine Agreement (FBA) and Federated Byzantine Fault Tolerance (FBFT) protocols. The FBA introduces the concept of a quorum slice, which is a subset of nodes that a given node chooses to trust. A quorum is a set, and each non-faulty member of it contains at least one quorum slice.
The UNL in RPCA is similar to a quorum slice. 
In Stellar, the ledger will not update the transaction until 100\% of nodes in a quorum slice agree, unlike in Ripple which requires only 80\% agreement. 
There are two mechanisms in the quorum slice model, federated voting and federated leader election. 
In voting, nodes vote on a statement and use a two-step protocol to confirm it. 
If each quorum of non-faulty nodes $v_1$ intersects each quorum of non-faulty nodes $v_2$ in at least one non-faulty node, then $v_1$ and $v_2$ are considered intertwined, and conflicting transactions will not be approved \cite{scp_2019}. 
In leader election, nodes pseudo-randomly select one or a small number of leaders in the quorum slice.

\begin{figure}[htp]
    \centering
    \includegraphics[width=\linewidth]{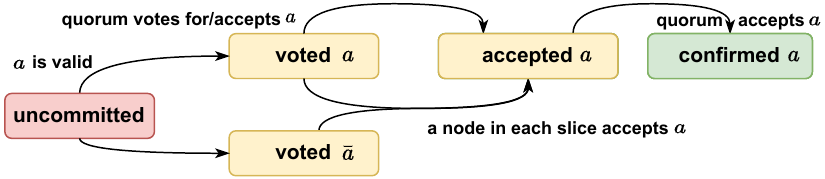}
    \caption{Federated voting process} 
    \label{fig:FBA}
\end{figure}

SCP is a global consensus protocol that includes three components: a nomination protocol, a ballot protocol, and a timeout mechanism. The nomination phase proposes new values as candidates for reaching an agreement using the statement \textsf{Nominate} $x$, where $x$ is a valid candidate consensus value. Each node that receives these values votes for a single value from among received ones. The nomination phase generates the same set of candidate values as a deterministic combination of all values on each intact node \cite{scp_2019}.
After the successful execution of the nomination phase, the nodes enter the ballot phase, which uses federated voting to commit or abort the values.
In FBA, as shown in Figure \ref{fig:FBA}, the three-step process involves a node broadcasting a valid statement $a$ and then accepting it if it doesn't conflict with any previously accepted values. If all members of the node's quorum set accept $a$, it is rebroadcasted. Finally, $a$ is confirmed if each node in the node's quorum accepts it, and the node confirms it as well.
However, there may be a \textsf{stuck} state since the node cannot conclude whether to abort or commit a value. SCP uses two statements \textsf{Prepare} and \textsf{Commit}, and a series of numbered ballots, to avoid stuck votes in the federated voting process. A statement $\textsf{Prepare} \langle n, x\rangle$ states that no value other than $x$ was or will ever be chosen in any ballot $\leq n$. Another statement $\textsf{Commit} \langle n, x\rangle$ states that value $x$ is chosen in ballot $n$. A node has to confirm the $\textsf{Prepare} \langle n, x\rangle$ statement before voting for the $\textsf{Commit} \langle n, x\rangle$ statement. Once \textsf{Commit} is confirmed, the value $x$ can be output by the node.
SCP provides liveness by using these two statements when the node determines a stuck ballot has been committed. The timeout mechanism is a crucial part of SCP. If the current ballot $n$ appears to be stuck, a new round of federated voting begins on a new ballot with a higher counter $n+1$. Unlike the Byzantine agreement, SCP allows participating nodes to determine their quorums, making it a significant difference. SCP utilizing FBA avoids stuck states and enables low latency and flexible trust.

%% file: 4_reliability.tex
\section{Reliability} \label{sec:reliability}
This section analyses the network reliability of consensus algorithms along the metrics of communication complexity, scalability, decentralization degree, and network models. Table \ref{tab:comparison} shows the comparison of above metrics in selected consensus algorithms.

\textbf{Complexity} refers to the number of messages required to reach a round of consensus. The smaller the complexity, the more efficient is the consensus algorithm. The number of messages in a normal case can differ from a leader failure situation. Table \ref{tab:comparison} presents the communication complexity of different protocols in normal situations and situations in which the leader fails. A message’s communication complexity affects the network’s scalability to some extent. If the complexity is higher, the nodes need more messages to accomplish consensus in one round. Therefore, algorithms in networks that accommodate more nodes tend to have lower complexity. In most consensus algorithms, the complexity is higher when a network leader fails since all participating nodes may require more rounds to have a new leader. The best case of complexity is linear for both normal and leader failure cases. In the category of PBFT and derivatives, PBFT's message complexity is $\mathcal{O}(n^2)$ when a non-malicious primary node operates without failure. Alternatively, it increases to $\mathcal{O}(n^3)$ if the primary node fails (processing view-change protocol). Since the leader is malicious, the protocol replaces it with a new leader through a view-change that includes at least $2f + 1$ signed messages. A new leader then broadcasts a new-view message containing proof of $2f + 1$ signed view-change messages. Validators will examine the new view-change message and broadcast it if it matches $2f+1$ view-change messages. Overall, the view-change complexity is $\mathcal{O}(n^3)$ while the leader fails. In this category, HotStuff can reduce the complexity to $\mathcal{O}(n)$ and guarantee responsiveness by using threshold signatures, three rounds of voting, and a chained structure to acknowledge a block \cite{Hotstuff_paper}. HotStuff can reach consensus pipelining without a complex view-change mechanism and improves efficiency. In the category of FBA, the complexity is determined by number of total nodes $n$ and number of nodes $K$ that formalized the federation. In normal case, complexity is $\mathcal{O}(nK)$. Upper bound is $\mathcal{O}(n^2)$ if and only if $K$ is close to $n$.

\textbf{Scalability} refers to the number of peering nodes that the algorithm can process and implies an upper bound on the size of the network. If a consensus protocol can support over 100 participants without losing network reliability, we conclude its scalability is \textsf{High}. A range of 20 to 100 is considered \textsf{Medium}, and anything lesser than 20 is considered to be \textsf{Low} suitability for scalability. Regardless of the message size and network environment, the scalability of a consortium blockchain network is also determined by the communication complexity. It can increase by reducing complexity. Therefore, a consortium blockchain atop a consensus algorithm with a lower complexity is always more scalable than a higher one. 
In the category of Paxos and derivatives, Paxos provides medium scalability, while Raft, its upgraded derivative, can support a relatively large network with high scalability.
In the category of PBFT and derivatives, the total number of broadcast messages grows quadratically with the increase in the total number of nodes, leading to rapid super-linear performance degradation. Therefore, PBFT is only suitable for consortium blockchain and private blockchain with low scalability.
In the category of FBA, SCP emphasizes maintaining the network's activity and allows any node to join each other's trust list for transactions if it follows the policy. With SCP, the Stellar network can run approximately 100 nodes \cite{Berger_Reiser_2018}. 

\input{table/T_assessment}

\textbf{Decentralization} implies that the existence of a relatively neutral entity functioning as a central node. In a round of reaching consensus, the node which decides the recording of transactions on the distributed ledger is considered as the central node \cite{Nguyen_Survey}. All other nodes keep the data consistent around it. In order to maintain the distributed state of the system, the role of each node (including central node) is subject to change. Therefore, we compare the degree of decentralization of the algorithm according to the recording node's selection rules and the number of selected recording nodes in each round. 
If all nodes are competitive while participating in consensus reaching and the probability of each node becoming the primary node is equal to each other, the degree of decentralization is defined as \textsf{High}. For instance, in Raft, every node has an equal chance of being elected by submitting a proposal and getting the votes from other nodes. 
If more than one node is selected as primary nodes or some nodes have higher priorities than other nodes, the degree of decentralization is assumed as \textsf{Medium}. 
The \textsf{Low} degree of decentralization only exists in cases where the primary nodes are pre-selected. i.e., the verification nodes in RPCA are pre–configured in the entire network. In a consortium blockchain, the degree of decentralization is flexible. It is unnecessary to require all nodes to have equal permission to participate in consensus reaching. To maintain the reliability of a consortium blockchain, some consensus algorithms use a low degree of decentralization as a tradeoff to reduce the number of validator nodes and seek a higher communication efficiency. Alternatively, the degree of decentralization can be decided by committee members as a business strategy.

The \textbf{network model} is an important metric that defines the ability of different message latencies to limit the blockchain network reliability. In general, there are three types of network models: (1) \textbf{Synchronous}: There is a known upper bound $\Delta$ on the latency of message communication between all nodes. The synchronous model offers an ideal communication pattern, which is hardly visible in real life but plays an important role in the theoretical study of distributed systems, and many early distributed consistency algorithms were designed under the assumption of synchronous networks. (2) \textbf{Asynchronous}: The above-mentioned upper bounds $\Delta$ does not exist, so the asynchronous model is more in line with the realistic Internet environment. Asynchronous is a more general case than synchronous. An algorithm that works for an asynchronous system can also be used for a synchronous system, but the converse does not hold. (3) \textbf{Partially Synchronous}: This model offers a communication pattern between the synchronous model and asynchronous model \cite{dwork1988consensus}. In the partially synchronous model, it is assumed that there is a globally stable clock GST (Global Stabilization Time), the message arrives in $\Delta$ time and the entire system may be in the asynchronous state before GST, but after GST, the whole system can return to a synchronous state. The timing assumption of the partial synchronization model is more in line with the real-world need for consensus algorithms, i.e., consensus can always be reached in a synchronized state; but once the network goes down, consensus may enter a period of blocking until it returns to normal. In addition, based on the relationships of the network models (synchronous $\subseteq$ partially synchronous $\subseteq$ asynchronous), if a consensus fully supports an asynchronous network, it theoretically supports synchronous and partially synchronous networks. Thus, a fully asynchronous consensus protocol provides high reliability.

%% file: table/T_assessment.tex
\setlength{\unitlength}{0.6em}
\newcommand\like[1]{\begin{picture}(1,1)
\ifnum0=#1\put(.5,.35){\circle{1}}\else
\ifnum10=#1\put(.5,.35){\circle*{1}}\else
\put(.5,.35){\circle{1}}\put(.5,.35){\circle*{.#1}}
\fi\fi\end{picture}}

\begin{table*}[!ht]
\caption{Consensus Algorithm Comparison}
\label{tab:comparison}
\begin{tabularx}{\textwidth}{@{}l*{15}{C}c@{}}
\toprule
Category $\rightarrow$ & \multicolumn{2}{|@{}c@{\hskip0in}|}{Paxos Derivatives} & \multicolumn{5}{@{}c@{\hskip0in}|}{PBFT Derivatives} & \multicolumn{2}{@{}c@{\hskip0in}|}{FBA} & \multicolumn{4}{@{}c@{\hskip0in}}{Randomized}\\ 
\midrule
\diagbox{Metrics}{Protocol} & \rotatebox[origin=c]{90}{Paxos \cite{Paxos_paper}} & \rotatebox[origin=c]{90}{Raft \cite{Raft_paper}}  & \rotatebox[origin=c]{90}{PBFT \cite{Pbft_paper}}  & \rotatebox[origin=c]{90}{RBFT \cite{Rbft_paper}}  & \rotatebox[origin=c]{90}{BFT-SMART \cite{Bftsmart_paper}}   & \rotatebox[origin=c]{90}{SBFT \cite{Sbft_paper}}  & \rotatebox[origin=c]{90}{HotStuff \cite{Hotstuff_paper} } & \rotatebox[origin=c]{90}{RPCA \cite{Rpca_paper}} & \rotatebox[origin=c]{90}{SCP \cite{SCP_paper}}  & \rotatebox[origin=c]{90}{PoET \cite{Poet_paper}} & \rotatebox[origin=c]{90}{Ouroboros \cite{ou_paper}} & \rotatebox[origin=c]{90}{HB-BFT \cite{miller2016honey}}  & \rotatebox[origin=c]{90}{Dumbo \cite{Dumbo_paper}} \\ \midrule
Complexity Nomal & $\mathcal{O}(n^2)$ & $\mathcal{O}(n)$ & $\mathcal{O}(n^2)$    & $\mathcal{O}(n^3)$  & $\mathcal{O}(n^2)$   & $\mathcal{O}(n)$  & $\mathcal{O}(n)$ &  $\mathcal{O}(nK)$  & $\mathcal{O}(nK)$ & $\mathcal{O}(n)$ & $\mathcal{O}(n)$ & $\mathcal{O}(n^3)$ & $\mathcal{O}(n^2)$ \\
Complexity Leader Failed$^{\dagger}$  & - & -  & $\mathcal{O}(n^3)$  & $\mathcal{O}(n^3)$ & $\mathcal{O}(n^3)$ & $\mathcal{O}(n^2)$ & $\mathcal{O}(n)$  &  $\mathcal{O}(nK)$ & $\mathcal{O}(nK)$ & - & - & - & - \\ 
Scalability & \like{5} & \like{8} & \like{1} & \like{1} & \like{5} & \like{5}  & \like{8} & \like{8} & \like{8} & \like{8} & \like{8} & \like{8}  & \like{8} \\ 
Decentralization & \like{8} & \like{8} & \like{1} & \like{5} & \like{8} & \like{8} & \like{8} & \like{1} & \like{8} & \like{8} & \like{8} & \like{8} & \like{8}\\ 
Network Model & P & P & P & P & P & P & P  & P  & P & S & S & A  & A \\ 
Latency & \like{1} & \like{1}  & \like{1}  & \like{1} &  \like{1}  & \like{1}  & \like{1} & \like{5} & \like{5} & \like{1} & \like{5} & \like{1}  & \like{1} \\ 
Throughput &  \like{5} & \like{5} & \like{1} & \like{8}  & \like{8}   & \like{8} & \like{8} & \like{5} & \like{1} & \like{1} & \like{1} & \like{8}  & \like{8} \\ 
Resource Consumption  & \like{1} & \like{1}  & \like{5} & \like{5} & \like{5}  & \like{1} & \like{1} & \like{1}  & \like{1} & \like{1} & \like{1} & \like{1} & \like{1} \\ 
Crash Fault Tolerance & 50\%  & 50\% & 33\% & 33\% & 33\% & 33\% & 33\%  & 20\% & 33\% & 50\% & 50\% & 33\%  & 33\%\\ 
Byzantine Fault Tolerance & - & - & 33\% & 33\% & 33\%   & 33\% & 33\% & 20\% & 33\% & - & - & 33\%  & 33\%\\
\bottomrule
\end{tabularx}
\begin{tablenotes}
\small
\item \like{8} $=$ \text{high}; \like{5} $=$ \text{medium}; \like{1} $=$ \text{low};
$\text{S}=\text{Synchronous}$; $\text{P}=\text{Partially Synchronous}$; $\text{A}=\text{Asynchronous}$; 
\item$^{\dagger}$ Leader Failed case is only suitable for PBFT derivatives and FBA. 
\end{tablenotes}
\end{table*}

%% file: 5_perf.tex
\section{Performance} \label{sec:performance}
This section provides an assessment of selected consensus algorithms regarding performance efficiency. In a consortium blockchain, domain services and transaction types vary. This creates a need for an efficient backbone for applications with low latency, high throughput, and low resource consumption. Table \ref{tab:comparison} shows evaluation results.

\textbf{Latency} is defined as the time elapsed from the moment a node submits a transaction to the time that the transaction is confirmed by blockchain. High latency means longer confirmation times for transactions, which can be an issue for applications that require faster processing times. On the other hand, low latency can provide faster confirmation times and improve the user experience. Thus, it is critical to consider the latency of a consensus algorithm when evaluating performance of blockchain. In this study, latency is classified as \textsf{High}, \textsf{Medium}, or \textsf{Low} \cite{Salimitari_Chatterjee_2019}. \textsf{High} latency is in the magnitude of minutes, \textsf{Medium} is in seconds, and \textsf{Low} is in milliseconds. 

\textbf{Throughput} refers to the block generation rate and the number of Transactions Per Second (TPS) the system can process. Block generation is expressed as time required for the entire process starting from the time when transactions are packaged into blocks up to the time when consensus is reached. TPS is determined by the size of block and block generation speed. TPS is calculated as the number of transactions in the block divided by the length of time required for the generation of the current block. We classify throughput into three categories. A protocol can provide $>$ 2,000 TPS is classified as a \textsf{High} throughput protocol. A TPS between 1,500 to 2,000 indicates \textsf{Medium} throughput, and a TPS $<$ 1,500 indicates a \textsf{Low} throughput. 
In the category of PBFT and derivatives, a blockchain implementation with BFT-SMART protocol by Symbiont can reach a throughput of 8000 TPS in a 4-node network cluster, which meets the expected performance of the original paper \cite{Bftsmart_paper}. 
In the category of FBA, the advantage of RPCA algorithm is its relatively high performance and efficiency. Ripple can generate a block every three seconds with a throughput that reaches 1500 TPS.

\textbf{Resource consumption} refers to the computing power, memory, input and output, and energy resources that each node consumes while reaching a consensus. Communication complexity is a theoretical proxy of resource consumption. Resource consumption is classified as \textsf{High}, \textsf{Medium}, or \textsf{Low} \cite{Salimitari_Chatterjee_2019}. 
\textsf{High} exists in some consensus algorithms that exhaust large resources for competing for a leader. This category only exists in public blockchains, for instance, when a node runs PoW it competes for a hash computation to get the transactions to be committed, including using a graphic card to acquire a high computing ability and thus is a drain on resources. 
Consensus algorithms defined as \textsf{Medium} in terms of their resource consumption usually are not a drain on resources when competing as a leader, but still need some due to a high communication complexity. 
\textsf{Low} resource consumption algorithms are ones that have an upper bound on communication complexity in any condition and do not require extensive computation to compete for a leader.

%% file: 6_security.tex
\section{Security} \label{sec:security}
This section provides an assessment of consensus algorithms regarding system security elaborated by fault tolerance. In terms of security, public blockchains face a high diversity of attack vectors. For instance, there are various attacks towards PoW, such as 51\% attack, eclipse attack, dust attack, empty block attack, selfish mining attack, block withholding attack, etc. \cite{Zhang_Xue_Liu_2019,saad2020exploring, anita2019blockchain,ye2018analysis,tosh2017security,sayeed2019assessing,conti2018survey,lunardi2022consensus}.However, the consensus protocols adopted in the consortium blockchains are more fault-tolerant and more controllable since they can detect and tolerate a certain number of malicious nodes. Therefore, consortium blockchains face relatively fewer security attacks and relatively low attack diversity. Therefore, only the impact of fault tolerance will be discussed in this section. The ability of a blockchain system to maintain uninterrupted operations when one or more of its components fail is intrinsic to its nature. Table \ref{tab:comparison} presents a comparison of selected consensus algorithms. 

\textbf{Fault Tolerance} is essential to security, and denotes the capacity of a distributed network to minimize the severity and frequency of network incidents, continue operations under stress, and recover as quickly as possible. We define security as the tolerance of the distributed network to malicious attacks or the amount of Byzantine behavior that the network can resist. Robustness is the ability of the network to maintain its performance in face of failures, or changes in topology or load. We consider two types of failures: Byzantine failures are those where some network participants may exhibit malicious behavior, such as sending conflicting messages, while crash failures are some participants are outage or unreachable. 

In the category of Paxos and derivatives, Paxos enables a distributed system to reach consensus when the number of non-failure nodes is greater than half of the total nodes. Raft was inspired by Paxos and its fault tolerance is very similar to Paxos. Neither Paxos nor Raft provide Byzantine fault tolerance. 
In the category of PBFT and derivatives, PBFT can tolerate both non-Byzantine errors and Byzantine errors, simultaneously, by sending broadcasts to the entire network in each round and allowing each node to participate in electing the primary node. This advanced mechanism ensures that PBFT has the capabilities to maintain consistency, ensure availability, and enable anti-fraud. 
RBFT was proposed for better resiliency during the process of Byzantine fault tolerance. In earlier BFT algorithms such as PBFT, Prime \cite{Amir_Coan_Kirsch_Lane_2011}, Aardvark \cite{clement2009making}, and Spinning \cite{Veronese_Correia_Bessani_Lung_2009}, if the primary node is malicious, the whole system's performance is degraded. 
The RBFT model proposes executing multiple PBFT protocol instances in parallel using multi-core machines, which can easily detect malicious nodes and prevent performance degradation.
If one or more Byzantine faulty nodes exist in the blockchain network, it has been shown that the maximum performance degradation of RBFT is 3\%, which is better than other protocols; for instance, Prime is 80\%, Aardvark is 87\%, and Spinning is 99\% \cite{Rbft_paper}.
HotStuff's responsiveness enables nodes to quickly confirm blocks under normal network conditions and can wait longer to confirm under limited network conditions. Overall, the algorithms in this category have the same level of fault tolerance on both crash and Byzantine faults.
In the category of FBA, the fault tolerance of SCP is the same lever as PBFT family. However, the fault tolerance of RPCA is lower than that of SCP and PBFT-like algorithms since the verification node is pre-configured, and the fault tolerance is bound to the number of verification nodes.   
In the category of Randomized, PoET and Ouroboros provide the same ability to tolerate crash faults. HB-BFT and Dumbo have the same level of fault tolerance with PBFT and derivatives on both crash and Byzantine faults.



%% file: 7_conclusion.tex
\section{Concluding Remarks}  \label{sec:conclusion}
Our intent for this paper was to carry out groundwork that informs researchers, developers, and the blockchain community at large of the current landscape of consensus methodologies and technologies and remaining challenges. The choice of a consensus algorithm has an enormous impact on the performance of a blockchain application. Therefore, ongoing research into the design and implementation of consensus algorithms will advance and facilitate the adoption of blockchain for diverse applications. Regardless of the type of blockchain used and its applications, a consensus algorithm is pivotal to the blockchain operation and must therefore be carefully designed. The primary research challenges that need to be addressed in the consensus mechanism domain for consortium blockchain are: (1) \textbf{Scalability enhancement}: While the consortium blockchain offers limited membership, the issue of scalability in a consortium blockchain is critical. As we discussed, the size of a network has implications for parameters such as fault tolerance that impact the blockchain’s efficiency. As business needs grow, the number of access nodes required by the platform may increase to keep pace with the platform’s expansion. Proactive approaches to building consortium blockchains that adapt to changing business and platform expansion needs must be considered to strengthen scalability. (2) \textbf{Algorithmic melding}: As applications and platforms evolve, consensus algorithms may require more flexibility in adapting to changing environments. The evolution of applications and platforms may introduce logical, but intricate, requirements for fusion between algorithms. Therefore, integrating different types of consensus mechanism algorithms in the future poses a distinct challenge to interoperability. (3) \textbf{Privacy-preservation}: The consortium blockchain needs authentication for participating nodes, which reduces the probability of possible attacks, to a certain extent. Nevertheless, we still need to consider the security and privacy of data on the consortium chain. The use of cryptography to ensure security and privacy of data on the blockchain while still conforming to the central paradigm of blockchain decentralization will be a tradeoff to consider. (4) \textbf{Performance improvement}: The potential to further several performance factors, such as increase in throughput, reduction in latency, and reduction in computational requirements for consensus algorithms must be considered. Each of these factors impacts the scalability of the blockchain. Therefore, ensuring increasing performance while reducing the impact on scalability is a challenge. (5) \textbf{Searching and storing optimization}: While the original philosophy of blockchain called for implementations to build a distributed ledger, the expectations for blockchain networks have evolved into data retrieval over the years. In this use scenario, a blockchain ledger is more like a distributed database without the capability of deleting and updating operations due to the immutability property of blockchain. Therefore, the consensus mechanisms that are built for blockchain should also consider whether the data storing and searching can be optimized accordingly. These challenges broadly identify the various areas of improvement for consortium blockchain algorithms. However, since these protocols are still under development and the applications leveraging these algorithms are continuously being refined, the scope of challenges for consensus algorithms in consortium blockchain will continue to be a work in progress.